\documentclass[english,showpacs]{revtex4}
\usepackage[T1]{fontenc}
\usepackage[latin1]{inputenc}
\usepackage{graphicx}

\makeatletter

\newcommand{\be}{\begin{eqnarray}}
\newcommand{\ee}{\end{eqnarray}}

\newcommand{\bea}{\begin{eqnarray}}
\newcommand{\eea}{\end{eqnarray}}

\usepackage{babel}
\makeatother

\begin{document}

\title{Top-pions and single top production at HERA and THERA}
\author{F.~Larios$^a$, F.~Pe\~nu\~nuri$^a$, M.~A.~P\'erez$^b$\\
{\small\it 
 $^a$Departamento de F\'{\i}sica Aplicada,
CINVESTAV-M\'erida, A.P. 73,  97310 M\'erida, Yucat\'an,
M\'exico}\\
{\small\it $^b$Departamento de F\'{\i}sica, CINVESTAV,
Apdo. Postal 14-740, 07000 M\'exico}\\
}
\noaffiliation 
\begin{abstract}
We study single top quark production at the HERA and THERA colliders
as coming from the FCNC vertices tc$\gamma$ and tcZ that appear at
one loop level in Topcolor-assisted technicolor models (TC2).
In contrast with previous expectations, we find that
the production cross section is of order $10^{-6}$ pb for HERA and
$10^{-4}$ pb for THERA (even lower that the SM prediction). Therefore,
none of the two colliders are capable to probe the pseudoscalar top-pion
or the scalar top-higgs predicted in TC2 models via single top production.
\end{abstract}

\pacs{14.80.-j;14.65.Ha;12.60.Nz}
\maketitle

\vspace{0.5cm}
\noindent{\large \bf 1. Introduction}
\vspace{0.5cm}

There has been an increased interest in studying forbidden or highly
suppresed processes as they become an ideal site to search for new
physics lying beyond the Standard Model (SM).
In particular, it is expected that if there is
any new physics associated with the mas generation mechanism,
it may be more apparent in the top quark interactions than
in the all the lighter fermions\cite{topreview}.  Along this line,
it has been suggested that t-channel single top production could
be rather sensitive to the non-SM flavor changing neutral
coupling (FCNC) vertices $tc\gamma$(Z)\cite{belyaev,peccei}.
While in general, vector and scalar FCNC vertices are very
suppresed by the GIM mechanism in the SM, some extensions
of the SM can generate some of them at the tree level\cite{rosado}.
This is the case of top-color assisted technicolor (TC2), where
a composite $t\bar t$ scalar (and a pseudo-scalar) appears as a
result of a strong interaction and can give rise to sizeable
FCNC $tch^0_t$ ($tc\pi^0_t$) couplings\cite{hill}.
Then, at the one loop level, these couplings can give rise to
effective $tcV$ couplings ($V=\gamma , Z$).

In particular, the pseudo-scalar flavor diagonal and the
flavor changing couplings that are relevant for the TC2
prediction of single top quark production are\cite{hill,yuan}:
\be \label{tc2coups}
\frac{m_t (1-\epsilon)}{\sqrt{2} F_t}
\frac{\sqrt{v^2_W-F^2_t}}{v_W} \;\; i \bar t \gamma^5 t \pi^0_t 
\nonumber \\
\frac{m_t}{\sqrt{2} F_t} \frac{\sqrt{v^2_W-F^2_t}}{v_W}
\; i k^{tc}_{UR} k^{tt*}_{UL} \;\; \bar t_L c_R \pi^0_t \, ,
\ee
where $F_t\sim 50$GeV is the top-pion decay constant, $v_W=v/\sqrt{2}$,
and the $k_U$ terms come from the diagonalization of the up and down
quark matrices.  Their numerical values are $k^{tt}_{UL}\sim 1$ and
$k^{tc}_{UR}\leq \sqrt{2\epsilon -\epsilon^2}$, with
$0.03 \leq \epsilon \leq 0.1$ a free parameter.  There are also
couplings involving the scalar top-higgs $h^0_t$, and their
effect on the production of single top at HERA is similar to
that of the top-pion.  In order to simplify
our discussion we will not consider them in this work.  
Notice that in this model the flavor changing scalar coupling only
involves the right handed component of the charm quark\cite{yuan}.

Recently, it was suggested that the neutral pseudo-scalar top-pion 
$\pi^0_t$ (or the neutral scalar top-higgs $h^0_t$) contribution to
the single top quark production cross section $\sigma (ec\to et)$
could be significant, of order 1-6 pb, at the HERA and THERA colliders
for a top-pion mass $m_\pi$ between 200 and 400 GeV\cite{yue}.

In the present letter, we re-examine this possibility and we find
that the cross section due to the top-pion is rather of order much
less than 1 fb.  In fact, the contribution of the top-pion (or
the top-higgs) is much smaller the SM contribution which is of
order less than 1 fb and is not observable at HERA \cite{baur,fritz}.
Unfortunately, none of these two colliders HERA and THERA are
thus capable to probe the TC2 scalar particles through single top
production.

\vspace{0.8cm}
\noindent{\large \bf 2. The FCNC single top production at HERA}
\vspace{0.5cm}

We consider the tree level FCNC process
\be
e(K_1) \;+\; c(P_1)\; \to \; e(K_2) \;+\; t(P_2)
\ee
where momentum conservation dictates $K_1+P_1=K_2+P_2$,
and the Mandelstam variables are defined as
$\hat s=(K_1+P_1)^2$, $t=(P_2-P_1)^2$ and $u=(P_2-K_1)^2$.
In this work, we take the top quark mass $m_t=175$GeV
and the charm quark as massless.  We also neglect terms
proportional to the electron mass $m_e$ in the differential
cross section.
However, we will use $m_e=0.5$MeV for the upper and lower limits
of the Mandelstam variable t in the phase space integral. 
This is done in order to avoid the divergence that appears
when the exchanged photon goes on-shell ($t=q^2=0$)\cite{fritz}.
The lower and upper limits for $\hat s = x S$ are
$m_t^2 \leq \hat s \leq S$.  
$S$ is the C.M. energy of the collider: $S=320$GeV for HERA
and $S=1$TeV for THERA.

Through a series of loop diagrams the FCNC tc$\pi^0_t$ vertex
gives rise to effective tcV vertices, with  V=$\gamma$,Z. 
These diagrams have been calculated and their results given in
terms of the following general effective tcV couplings\cite{yue}:
\be \label{yuecoups}
\Lambda^\mu_{V\bar t c} =
ie \left( \gamma^\mu
(F_{1V}+F_{2V} \gamma^5)
- P_2^\mu (F_{3V}+F_{4V} \gamma^5)
+ P_1^\mu (F_{5V}+F_{6V} \gamma^5) \right).
\ee
(In the notation of \cite{yue} $P_1=P_c$ and $P_2=-P_t$.)
The $F$ coefficients are given in Ref.\cite{yue}.  We have
computed them and have found agreement, except for the fact
that the left handed component of the c quark cannot
participate in the tc$\gamma$ vertex.  The reason for this is
that it is only the right handed component of the c quark that
appears in the tc$\pi^0_t$ vertex, which gives rise to tc$\gamma$.
Nevertheless, this correction does not change in any way the
fact that the cross section for single top production turns out
to be negligible for our TC2 model.

\vspace{0.8cm}
\noindent{\large \bf 
3. General tc$\gamma$ vertex with current conservation}
\vspace{0.5cm}

For this t-channel process it is well known that HERA will
get the bulk of the production cross section from the region
where the momentum transfer is very small and the
exchanged photon is quasi-real\cite{fritz}.  The contribution
from Z exchange is several orders of magnitude smaller.
In fact, the experimental situation is such that the scattered
positron usually escapes through the rear beam pipe, and events
with $Q^2$ greater than 1GeV$^2$ are rejected\cite{zeus}.
The reason for this is because the photon propagator tends
to infinity in the limit $q^2\to 0$, and the phase space
integration must be done carefuly.  In particular, the effective
$tc\gamma$ couplings of Eq.~(\ref{yuecoups}) are not the most
convenient to work with.  It turns out that the coefficient $F_{1}$
is different from zero in the on-shell limit $t=q^2=0$
(see Eq.~(\ref{fnums1}) below).
This means that if taken isolated (disregarding the interference
with $F_3$ and $F_5$) its contribution to the cross section
diverges very rapidly.  Indeed, the contribution of the other
coefficients $F_3$ and $F_5$ will also diverge, but in such a
way as to cancel out the contribution from $F_1$.
Unlike the strategy followed by Ref.~\cite{yue}, we will not use
the coupling in the form of Eq.~(\ref{yuecoups}).  Instead,
we will perform a Gordon decomposition to transform it into a
more suitable form; one that has no such big cancellations.

With the bulk of the contribution coming from a quasi real photon
we should bear in mind that electromagnetic vector current
conservation implies that the coupling $\bar t \gamma_\mu c$
must vanish when the photon goes on shell\cite{deshpande}.
In order to make this apparent, we take the $tc\gamma$ coupling of
Eq.~(\ref{yuecoups}) and change it from the ($\gamma^\mu$,
$P_1^\mu$, $P_2^\mu$) basis to the ($\gamma^\mu$,
$\sigma^{\mu \nu} q_\nu$, $q^\mu$) basis\cite{deshpande}.
We re-write Eq.~(\ref{yuecoups}) through a Gordon decomposition:
\be
\Lambda^\mu_{V\bar t c} =
ie \left( \gamma^\mu
(V_{V}-A_{V} \gamma^5)
+ i\sigma^{\mu\nu} q_\nu (F_{7V} + F_{8V} \gamma^5)
+ q^\mu (F_{9V}+F_{10V} \gamma^5) \right)
\label{gordon}
\ee
where $q=P_2-P_1$ is the momentum of the exchange photon (or Z),
and with the new coefficients given by
\be
V &=& F_1 + m_t \frac{F_5-F_3}{2}  \nonumber \\
A &=& F_2 + m_t \frac{F_6-F_4}{2} \nonumber \\
F_7 &=& \frac{F_3-F_5}{2} \nonumber \\
F_8 &=& \frac{F_4-F_6}{2} \nonumber \\
F_9 &=& -\frac{F_3+F_5}{2} \nonumber \\
F_{10} &=& -\frac{F_4+F_6}{2} \nonumber
\ee

Based on the results of Ref.\cite{yue} we can make a numerical
calculation of the coefficients defined above.  First, let us see
the values of $F_{3-5}$ for a top-pion mass $m_\pi = 200$GeV and
for $q^2=0$:
\be
F_{1\gamma} &=& -6.918 \times 10^{-3} f(\epsilon)
\label{fnums1} \\
m_t F_{3\gamma} &=& -4.1198 \times 10^{-3} f(\epsilon)
\nonumber \\
m_t F_{5\gamma} &=&  9.7162 \times 10^{-3} f(\epsilon).
\nonumber 
\ee
Here, we have separated the factor
$f(\epsilon)=(1-\epsilon)\sqrt{2\epsilon-\epsilon^2}$
that gives the dependence on $\epsilon$.
The cancellation in Eq.~(\ref{gordon}) can be easily verified.
For $q^2$ small but not zero we have that $V_{\gamma (Z)}$ varies
linearly with $t=q^2$.  Nevertheless, all the other form
factors remain nearly constant for small and even
greater (than 1 GeV$^2$) values of $t$.
As mentioned before, the fact that $V_\gamma$ is directly
proportional to $q^2$ is expected from electromagnetic current
conservation\cite{deshpande}.  To illustrate this point and to
see that the dominant contribuion comes from the magnetic
transition coefficient $F_{7\gamma}$, let us now note the
numerical values of $V_\gamma$, $F_{7\gamma}$ and $F_{9\gamma}$
for a top-pion mass $m_\pi = 200$GeV and for $-q^2 \le 1$GeV$^2$:
\be
V_\gamma &=& 1.6 \times 10^{-3} \times \frac{t}{m_t^2}
\;\; \times f(\epsilon) \label{fnums2} \\
m_t F_{7\gamma} &=& -6.9 \times 10^{-3}
\;\; \times f(\epsilon) \nonumber \\
m_t F_{9\gamma} &=& -2.8 \times 10^{-3} f(\epsilon)
\nonumber 
\ee
Because $V_\gamma$ is proportional to $t=q^2$ its contribution
to the cross section will not diverge in the $q^2=0$ limit.
This is not quite the case for $F_7$ and $F_9$; for them there is
still a divergent behaviour, although this time it is only a
logarithmic one.  On the other hand, the $F_9\; \bar t q^\mu c$
coupling contribution is proportional to the electron mass, and
it is therefore much smaller than that of $F_7$.

\begin{figure}
\includegraphics[width=1.0\columnwidth]{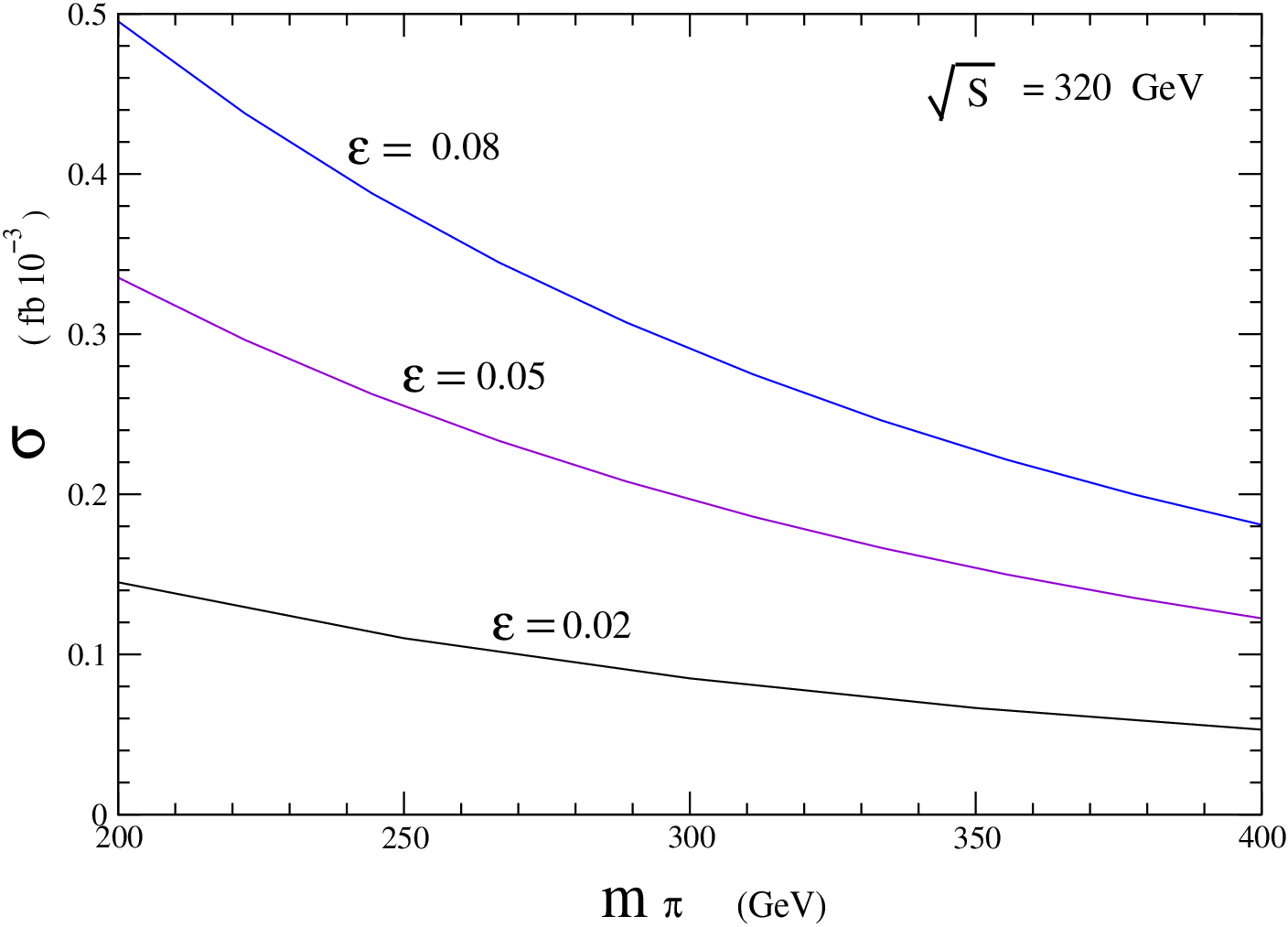}
\caption{\label{hera} Single top production cross section from $\pi^0_t$
as a function of $m_\pi$ for the HERA collider.}
\end{figure}

\begin{figure}
\includegraphics[width=1.0\columnwidth]{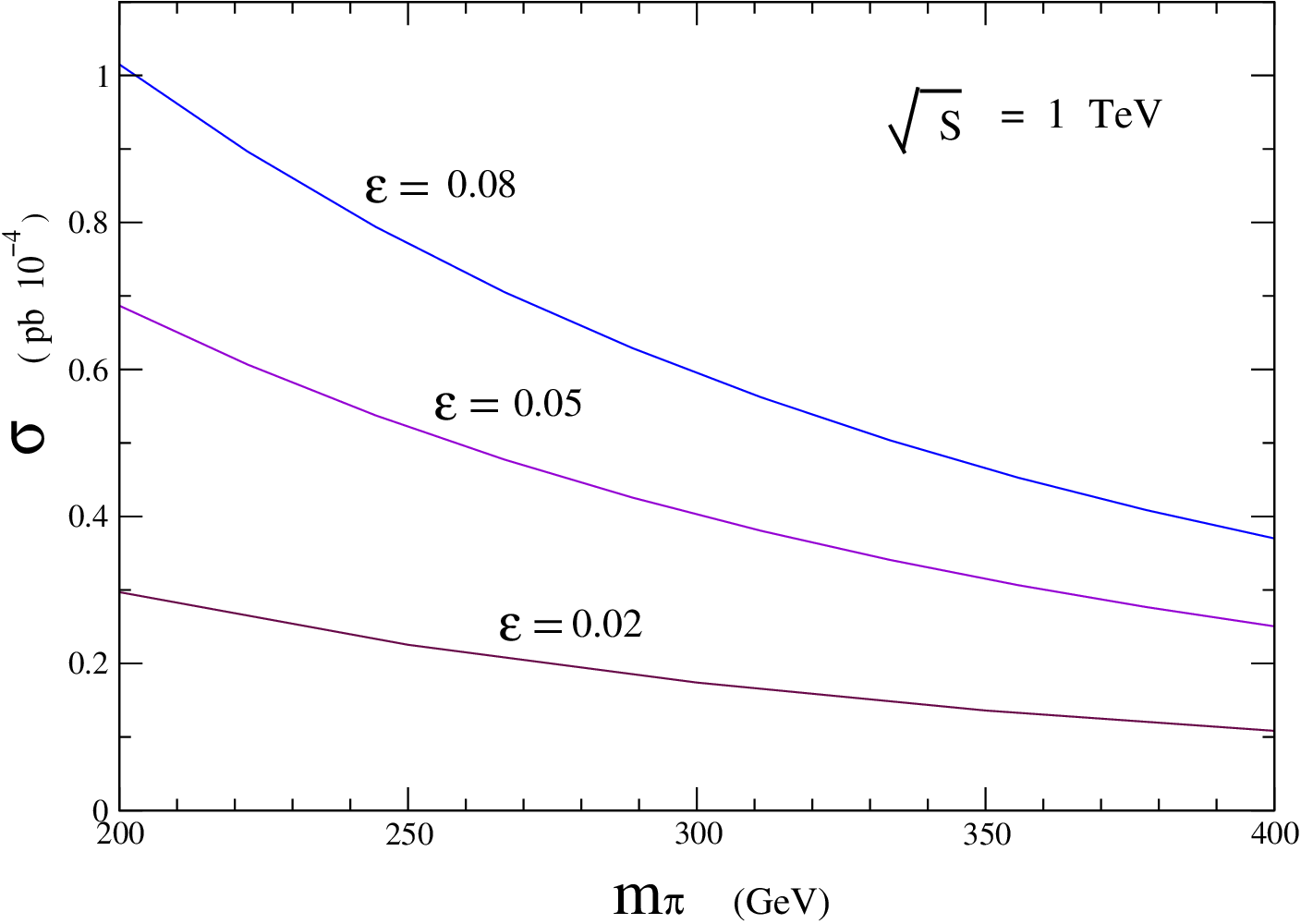}
\caption{\label{thera} Single top production cross section from $\pi^0_t$
as a function of $m_\pi$ for the THERA collider.}
\end{figure}

The coefficients for tcZ are of similar values, even though
their contribution is very small we have included them in our
numerical results shown in Figs.~(\ref{hera}) and (\ref{thera}).

\vspace{0.8cm}
\noindent{\large \bf 4. Discussion and results}
\vspace{0.5cm}

The differential cross section contribution from photon exchange,
disregarding terms proportional to the electron mass,
is given by\cite{kidonakis}:
\be
\frac{d\sigma}{dt} &=& \frac{2\pi \alpha^2}{\hat s^2}
\frac{M_\gamma}{t^2} \\
M_\gamma &=& F^2_{7\gamma} \;t\; \left[ \;
2\hat s (m^2_t-\hat s) - m^4_t -t(2\hat s-m^2_t)
\; \right] \nonumber
\ee
As mentioned in Ref.\cite{kidonakis} the bulk of the cross section
is given by the photon exchange diagram ($M_\gamma$).  This is
because of its logarithmic divergent behavior that is taken under
control with the electron mass.  However, our numerical results
include the (negligible) contribution from Z exchange.

The total cross section is given by:
\be
\sigma &=& \int^1_{x_{min}} dx \; \phi (x)\;\;
\int^{t_{max}}_{t_{min}} dt \frac{d\sigma}{dt}
\nonumber
\ee
with $\phi (x)$ the charm quark PDF and $x_{min}=(m_t+m_e)^2/S$
the minimum value of $x$.    The C.M. energy of the collider
is $S=320$GeV for HERA and $S=1$TeV for THERA.

The limits for $t=-Q^2$ are:
\be
t_{min(max)} &=& \frac{m^4_t}{4\hat s}-(k_{1cm}
\pm k_{2cm})^2 \label{tlimits} \\
k_{1cm} &=& \sqrt{\frac{(\hat s + m^2_e)^2}{4\hat s}-m^2_e}
\nonumber \\
k_{2cm} &=& \sqrt{\frac{(\hat s + m^2_e - m^2_t)^2}{4\hat s}-
m^2_e}
\nonumber
\ee

We have evaluated the total cross section with the
CTEQ6M PDF\cite{cteq}, running at a fixed scale $\mu = m_t$.
We have also run $\phi (x)$ with the energy scale and have
seen no significant change.

In Figure~(\ref{hera}) we show the production cross section for the
process $ec\to et$ coming from the FCNC vertices tc$\gamma$ and
tcZ.  For a $m_\pi = 200$GeV and $\epsilon = 0.08$ we obtain
a tiny cross section $\sigma = 0.5 \times 10^{-6}$ pb, which is
even smaller than the SM contribution to single top production.
As mentioned before, the dominant contribution comes from the photon
diagram so that $\sigma$ is proportional to $F^2_{7\gamma}$.
We can compare with Ref~\cite{kidonakis}, when they take
$\kappa_\gamma = 0.1$ for $ec\to et$ at HERA they obtain a very
small cross section of order $0.002$pb including NNLO QCD corrections.
They instead prefer to discuss $eu\to et$ because of the small sea
charm quark pdf density.  In our model, the tu$\pi^0_t$ coupling is
very small and it will not generate a sizeable tu$\gamma$ effective
vertex.  Also, notice that in our notation we have
$m_t F_{7\gamma}= 2\kappa_\gamma$.
In Fig~(\ref{hera}) we see that for $\epsilon =0.08$
and $m_\pi=200$GeV the cross section is $0.5\times 10^{-6}$pb.
This is four orders of magnitude smaller than the cross section
of Ref.~\cite{kidonakis} because in our case
$m_t F_{7\gamma}= 0.0025$ which corresponds to $\kappa_\gamma =0.001$
two orders of magnitude smaller.  This is only a rough comparison,
as we do not include NNLO QCD corrections.

In figure~(\ref{thera}) we show the production cross section for
the THERA collider.  It is 4-5 orders of magnitude greater than
the one obtained at HERA, but still too small for an experiment
with estimated luminosity of 500 pb$^{-1}$.

As a final remark, let us compare with the results of Ref.~\cite{yue}.
There, the cross section values are 6 orders of magnitude greater.
As mentioned before, the the cross section in Ref.~\cite{yue} was
calculated from an effective coupling in the form of
Eq.~(\ref{yuecoups}).  This means that a
very large cancellation must take place in the small $q^2$ region,
and any standard numerical integration may not be able to handle
this situation properly.  Rather, the numerical integration is
likely to become unstable for small $t$.
Perhaps this is the reason why a (very large) cut in
$t_{min}$($=-0.001$GeV$^2$) had to be applied in their calculation.
In this work, we have instead used the effective coupling in the
form of Eq.~(\ref{gordon}).  This new coupling does not induce the
large cancellations of the previous one.  The numerical integration
is stable, even in the low $t=q^2$ region.  We use the limits
given in Eq.~(\ref{tlimits}), for which $t$ can reach small values
of order $t=-10^{-8}$ GeV$^2$.

In conclusion, we have shown that neither HERA nor THERA can probe the
effects of a top-pion (or top-higgs) of TC2 models via single top quark
production.

\acknowledgments
We want to thank C.-P. Yuan for helpful discussions.
We thank Conacyt for support.

\end{document}